\documentstyle[twoside,fleqn,espcrc2,epsfig]{article}

\newcommand{\beq}{\begin{equation}}
\newcommand{\eeq}{\end{equation}}
\newcommand{\bea}{\begin{eqnarray}}
\newcommand{\eea}{\end{eqnarray}}

\newcommand{\CZ}{{\cal Z}}

\newcommand{\CG}{{\cal G}}

\newcommand{\gstr}{\gamma_{str}}
\newcommand{\pa}{\partial}

\newcommand{\nn}{\nonumber}

\title{Three-state Complex Valued Spins Coupled to Binary Branched Polymers in 
Two-Dimensional Quantum Gravity}
\author{Jo\~ao D. Correia\address{Department of Physics, Theoretical Physics, 
        Oxford University \\ 
        1 Keble Rd., Oxford OX1 3NP, U.K.}%
        \thanks{Work partially supported by Praxis XXI grant BD2932/94. 
Presented at the conference by J. D. Correia.}
        and 
 John F. Wheater\address{Department of Physics, Theoretical Physics, 
        Oxford University \\ 
        1 Keble Rd., Oxford OX1 3NP, U.K.}}
       
\begin{document}
\begin{abstract}
A model of complex spins (corresponding to a non-minimal model in the language 
of CFT) coupled to the binary branched polymer sector of quantum gravity
is 
considered. We show that this leads to new behaviour. 
\end{abstract}

\maketitle

\section{Introduction}

Branched polymers play an important role in the discretized version of 
bosonic non-critical strings. Both numerical simulations \cite{num} and
semi-rigorous results (\cite{me}, \cite{harris}, \cite{david}) lead us to 
believe they
dominate the theory for values of the central charge $c$ greater than 1.
These configurations are caracterized by a critical exponent $\gstr=\frac{1}{2}$.
It is possible that coupling non-minimal models to branched polymers
might lead to new behaviour \cite{stau}. We investigate this possibility.
We also consider the interpolation between a fixed configuration and the
fully fluctuating case.

\section{Binary Trees}

The properties of the ensemble of tree graphs are well know \cite{kaz}.
In this paper we will consider trees made of cubic vertices (a subset of
the full branched polymer ensemble) but modify the 
graphs slightly so that all the external lines except the root are
attached to another line. 
Letting $T_N$ be the number of graphs with $N$
external vertices (not counting the root) we have
\beq T_N = \sum_{k=1}^{N-1} T_{N-k} T_k \mbox{\ , $T_1=1$}
\label{recurr} \eeq
or
\beq T(z) = \sum_{N=1}^{\infty} z^N T_N = \frac{1}{2} \left ( 1-
\sqrt{1-4 z} \right ) \label{gen} \eeq
The exponent $\gstr$ for the ensemble of graphs with one marked point
(the root in this case) is defined so that the generating function $\CG$
for the number of graphs of a given size has leading non-analytic
behaviour
\beq \frac{\pa \CG}{\pa z} = (z_{cr}-z)^{-\gstr} \label{leading} \eeq
so for the tree ensemble $\gstr=\frac{1}{2}$, the same value as the
full branched polymer case.

\section{Matter coupled to binary trees}

\begin{figure}[h]
\vglue 3mm
\epsfig{figure=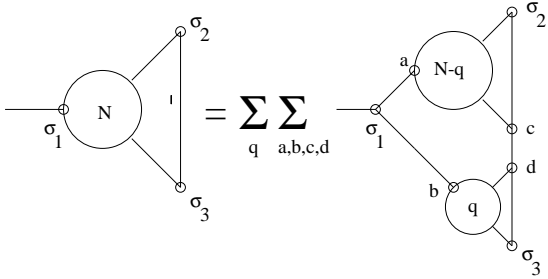, angle=0,width=72mm}
\vglue -1mm
\caption{The recurrence relation for spins coupled to binary trees.}
\label{iter2}
\end{figure}

We can extend the model by coupling matter to the trees by placing an
Ising spin $\sigma_{i} = \pm 1$ at each of the vertices. We obtain the 
recurrence relation (see fig. \ref{iter2})
\bea Z_{N} &=& \frac{1}{2^{4}} \sum_{q=1}^{N-1} \sum_{abcd} (1+t \sigma_{1}
a) (1+t \sigma_{1} c) (1+ t b d) \nn \\ 
&\ & Z_{N-q} (a, \sigma_{2}, b) Z_{q} (c, d, \sigma_{3}) \label{z2} \eea
In the absence of a magnetic field the dependence of $Z_N$ on the external
configuration $\sigma = (\sigma_{1}, \sigma_{2}, \sigma_{3})$ must take
the form
\beq Z_{N} (\sigma) = A_{N} + \sigma_{1} \sigma_{2} B_{N} + \sigma_{1}
\sigma_{3} B_{N} + \sigma_{2} \sigma_{3} C_{N} \label{ansatz2} \eeq
Inserting (\ref{ansatz2}) into the partition function
(\ref{z2})  and equating coefficients depending on the same combination of
spins on both LHS and RHS, we find a system of coupled equations for the
$A_{N}$,
$B_{N}$, $C_{N}$ in terms of $A_{N-1}$, $A_{N-2}$, etc. Defining the
grand-canonical partition functions 
\beq A (z,t) = \sum_{N=1}^{\infty} z^{N} A_{N} (t) \eeq
and similarly for $B (z,t)$, $C(z,t)$, we obtain the system of equations: 
\bea A &=& z + A^{2} + t^{3} B^{2} \label{z2in} \\
     B &=& z + t A B + t^2 B C \label{z2mid} \\
     C &=& z + t C^2 + t^2 B^2 \label{z2fin} \eea

Singularities in any of the functions $A$, $B$ or $C$ (or in any of their
derivatives) will signal a critical point in the full partition function
(\ref{z2}). It can be shown that one has $\gstr=\frac{1}{2}$ for
$t=[0 \mbox{,}1]$.The critical structure 
of the theory is not changed by the addition of Ising spins. This result
is expected because the Ising spins never have a diverging correlation
length in less than two dimensions and so cannot affect the global
properties of the geometry. Again we recover the behaviour observed when
we couple the Ising model to the full branched polymer ensemble.

\begin{figure}[h]
\vglue 3mm
\epsfig{figure=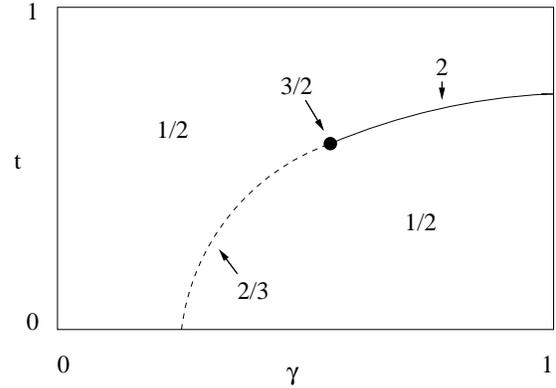, angle=0,width=72mm}
\vglue -1mm
\caption{The phase diagram for complex valued spin coupled to binary trees;
here $t=\frac{1}{2} (\mu + \nu)$ and $\gamma=\frac{1}{2} (\mu - \nu) (\mu + \nu
-2)^{-1}$.}
\label{phased1}
\end{figure}

We now consider a generalised Ising model in which the  spins take the values
$1$, $ e^{\pm \frac{2 \pi i}{3}}$. 
By allowing the partition function to include complex weights, 
we obtain a richer phase structure than that of the 2-state case. 
The partition function is given by
\beq {\CZ} = \sum_{trees} \sum_{\{S_{i}\}} \prod_{links} 
(1 + \mu S_{i} S_{j}^{\dagger} + \nu S_{i}^{\dagger} S_{j}) \eeq
Again we construct the most general form for the partition function:
\begin{eqnarray} Z_{N} &=& A_{N} + S_{1} S_{2}^{\dagger} 
B_{N} + S_{1} S_{3}^{\dagger} \tilde B_{N}  + S_{1}^{\dagger} 
S_{2} D_{N} \nn \\ &+& 
S_{1}^{\dagger} S_{3} \tilde D_{N} +  S_{2} S_{3}^{\dagger} C_{N} + 
S_{3} S_{2}^{\dagger} E_{N} \nn \\  &+& S_{1} S_{2} S_{3} F_{N} + S_{1}^{\dagger} 
S_{2}^{\dagger} S_{3}^{\dagger} G_{N} \label{A} \end{eqnarray}
The system of equations which emerges from this is rather more complicated
than the obtained from the simple Ising (\ref{z2in})-(\ref{z2fin}). But
it still falls under the same generic category
and can be solved for different values of the coupling constants
$\mu$ and $\nu$; the results are neatly summarized in the phase diagram
of the model, fig. \ref{phased1}.

We see that, unlike the case of the Ising, the non-minimal ${\bf Z}_3$
model leads to the possibility of critical exponents other than
$\gstr=\frac{1}{2}$.

Now we want to modify the
problem so that one particular sort of graph is picked out and given a
different weight; this is the ladder graph. For a given $N \geq 3$ there 
are precisely two of these; However, $N=1$ and $N=2$ are
special because there are only two graphs in the whole ensemble. All the
non-ladder graphs we will call ``trees''; so in fact there are no trees
for $N<4$. We will begin by considering the case of pure gravity.

Letting $L_N$ be the number of ladders with $N$ external vertices we
have
\bea L_1 = L_2 = 1 \nn \\
L_{N \geq 3} = 2 \nn \eea
so that
\beq L(z) = \frac{2 z}{1-z} - z - z^2 = z \left ( \frac{1+z^2}{1-z}
\right ) \label{lz} \eeq
and so the exponent $\gstr$ takes the value $2$ for a pure ladder
ensemble.
The number of trees $T_N$ satisfies 
\bea T_1 = T_2 = T_3 = 0 \nn \eea
\bea T_N = &q& \left (  \sum_{k=1}^{N-1} (T_{N-k}+ L_{N-k})(T_k + L_k) \right ) \nn \\ 
&-&  \left ( \sum_{k=1}^{N-1} L_{N-1} - \delta_{N3} \right ) \label{tn} \eea
The factor $q$ enables us to assign a different relative weight in the
ensemble to trees and ladders: a typical tree with $N$ external vertices 
gets a factor of
$q^{N-1}$. For the generating function we find 
\beq T(z) = q \left \{ (T(z) + L(z))^2 -z L(z) -z^3 \right \} \label{tz}
\eeq
which is easily solved to yield the generating function ${\CG} = T(z) + L(z)$
 for the modified ensemble:
\beq {\CG}(z) =  \frac{1}{2q} \left ( 1- \sqrt{1-4 q \left (
(1-q z) L - q z^3 \right )} \right ) \label{gz} \eeq
For $q=1$ this just becomes the usual tree generating function
with $\gstr=\frac{1}{2}$ whereas at $q=0$ it is equal to
$L(z)$. However for any positive non-zero value of $q$ we find that
$\gstr=\frac{1}{2}$. This is easily seen by considering the behaviour of
the argument of the square root as $z$ is increased from zero; as $z$
increases $L(z)$ increases but before it diverges the argument of the
square root must vanish (because it goes to $-\infty$ if $L(z)$ goes to
$\infty$). Thus only at $q=0$ exactly do we manage to ``freeze out'' the
general trees and get a system which contains only ladders. This
behaviour is very reminiscent of the $R^2$ model discussed in \cite{r2}; 
this remains in the ordinary gravity phase for all finite
$R^2$ coupling.

A similar phenomenon continues to occur when we couple matter to the ensemble. 
For  $q=0$ we have an exceptional regime, where the exponent $\gstr$
takes values different from those of the dynamical phase, which sets in
for any positive, non-zero value of $q$.

\section{Conclusions}

We have shown how the introduction of complex matter can change the value of 
$\gstr$ for a branched polymer ensemble very much as the introduction of
ordinary matter can change it in two dimensional quantum gravity. In
this models we can also solve the problem of interpolation between a
``gravity'' regime and a ``crystalline'' regime; we have found that
crystalline behaviour is only obtained when geometry fluctuations are
completely forbiden.


\begin{thebibliography}{9}

\bibitem{num}{G. Thorleifsson, these Proceedings}


\bibitem{me}{J. D. Correia and J. F. Wheater, Phys. Lett. {\bf B388} (1996)
707-712}

\bibitem{harris}{M. G. Harris and J. F. Wheater, Nucl. Phys. {\bf B427} (1994) 
111-138}

\bibitem{david}{F. David, Nucl. Phys {\bf B487} (1997) 633-649}

\bibitem{stau}{M. Staudacher, Nucl. Phys. {\bf B 336} (1990) 349}

\bibitem{kaz}{D. V. Boulatov, V. A. Kazakov, I. K. Kostov and A. A. Migdal,
Nucl. Phys. {\bf B257} (1985) 433}

\bibitem{r2}{V.A. Kazakov, M. Staudacher and T. Wynter, Commun. Math.
Phys 177 (1996) 451, Nucl. Phys. {\bf B471} (1996) 309-333}


\end{thebibliography}
\end{document}